\documentclass[a4paper, epsf]{spie}  
\usepackage[]{graphicx}

\def\farcm@mss{\ensuremath{.\mkern-4mu{}^{\prime}}}%
\def\farcs@mss{\ensuremath{.\!\!{}^{\prime\prime}}}%

\title{High redshift galaxy surveys}
\author{Masanori Iye \\[12pt]
National Astronomical Observatory, Osawa, Mitaka, Tokyo 181-8588 JAPAN\\[12pt]
}

\authorinfo{Plenary talk, SPIE 2008 Astronomical Instrumentation, Marseille \\
\vspace{5mm} M.Iye: E-mail: m.iye@nao.ac.jp, Telephone: +81-422-34-3520, Fax: +81-422-34-3527}

\begin{document}
\maketitle

\begin{abstract}
A brief overview on the current status of the census of the early universe population is given.  Observational surveys of high redshift galaxies provide direct opportunities to witness the cosmic dawn and to have better understanding of how and when infant galaxies evolve into mature ones. It is a much more astronomical approach in contrast to the physical approach of to study the spatial fluctuation of cosmic microwave radiation.  Recent findings in these two areas greatly advanced our understanding of the early Universe.  I will describe the basic properties of several target objects we are looking for and the concrete methods astronomers are using to discover those objects in early Universe. My talk starts with Lyman alpha emitters and Lyman break galaxies, then introduces a clever approach to use gravitational lensing effect of clusters of galaxies to detect distant faint galaxies behind the clusters.  Finally I will touch on the status and prospects of surveys for quasars and gamma-ray bursts.
\end{abstract}


\section{INTRODUCTION}
Since the discovery of the expansion of the Universe by Edwin Hubble in 1929, astronomers with ever more powerful telescopes surveyed the sky to find more and more distant galaxies. By studying distant galaxies, one can look back the early history of the Universe. Partridge and Peebles$^1$, in their classical 1967 paper, predicted the properties of primordial galaxies and pointed out that these galaxies with redshifted Lyman alpha emission are the targets observational astronomers should look for.  Many attempts followed using 4m class telescopes for next three decades. This was, however, not an easy task$^2$. 

Astronomers of this decade developed various techniques to isolate distant objects; narrow band imaging surveys for Lyman alpha emitting galaxies$^{3-28}$, multi-band photometric surveys for Lyman break galaxies$^{29-38}$, searches for amplified images of gravitationally lensed galaxies$^{39-47}$, quasars$^{48-54}$ and studies of sporadic gamma ray bursts$^{55-57}$ in high redshift galaxies. Galaxies up to redshift $z=6.96^{18}$ were spectroscopically confirmed and there are additional candidate galaxies that appear to be at redshift $z>7^{34-37,,41,44,45}$. 

The current picture of the big bang Universe indicates that the expanding universe cooled rapidly to form neutral hydrogen from protons and electrons at 380,000 years after the big bang. This is the epoch when the photons are decoupled from the matter. The density fluctuation of the dark matter and the matter grew by gravitational interaction and it is conceived that the first generation of stars were born at around 200 million years after the big bang.  Initial set of formed stars contained wide range of mass spectrum. The absence of metal elements in the primordial gas helped to form massive stars. Due to the strong UV radiation from those newly formed massive hot stars, the surrounding intergalactic matter was gradually re-ionized. A kind of "Global Warming of the Universe".  When and how these re-ionization process took place is not observationally clarified yet but WMAP5 results$^{59}$ suggest $z\sim11$ if the re-ionization was an instantaneous event. It is more likely that the cosmic re-ionization could have taken place in an extended period sometime during $6 < z <17$. 

Detailed observations deep into the era beyond $z=7$ is, therefore, crucial.  Some of the recent number counts of galaxies at $5.7 < z < 7$ indicate significant decrease in the number density of Lyman alpha emitting galaxies$^{16-18}$, which could either be due to the evolution of galaxies possibly through merging processes or due to the increasing fraction of neutral hydrogen blocking Lyman-$\alpha$ emitting galaxies at high redshift. 
I will describe the target population of galaxies in the early Universe and the technique astronomers are employing to find those objects together with some recent results.

\section{NARROW BAND SURVEY FOR LYMAN ALPHA EMITTERS}
What are Lyman alpha emitters, that are often abbreviated ash LAEs?　They are thought to be star-forming young galaxies with star formation rate from 1 to 10 solar mass per year.   Hot massive stars produce strong UV radiation field and ionize the interstellar gas. The ionized hydrogen recombines and cools by emitting a Lyman alpha photon to settle down to the lowest ground level.  The amount of stars produced in these galaxies is not yet very large as the usual continuum radiation from stars is not necessarily conspicuous.  The spectra of LAEs are therefore characterized by strong Lyaman-$\alpha$ emission line as shown in Fig.1.

How to find those LAEs?   It would be natural to catch the Lyman alpha emission line signal from these galaxies.  Since these objects are so faint, one has to consider the properties of the sky background, actually foreground radiation from the Earth's atmosphere. The night sky glows ever brighter at longer wavelength.  In the wavelength region below 1 micron , where Si-CCDs are sensitive, the night sky spectrum shows strong bands of OH emission lines as shown in the lowest panel of Fig.2.  The gaps between these OH bands are nice dark windows to probe deep space.

\begin{figure}
\begin{center}
\begin{tabular}{c}
\includegraphics[height=5cm]{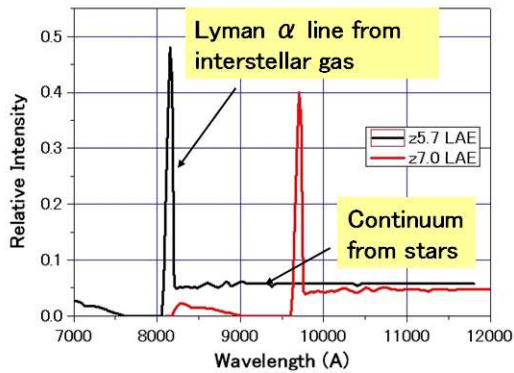}
   \end{tabular}
   \end{center}
\caption[fig1]
 { \label{fig:fig1}
Typical spectra of Lyaman-$\alpha$ emitters showing conspicuous Lyman alpha emission lines.}
\end{figure}

\begin{figure}
  \begin{center}
  \begin{tabular}{c}
    \includegraphics[height=6cm]{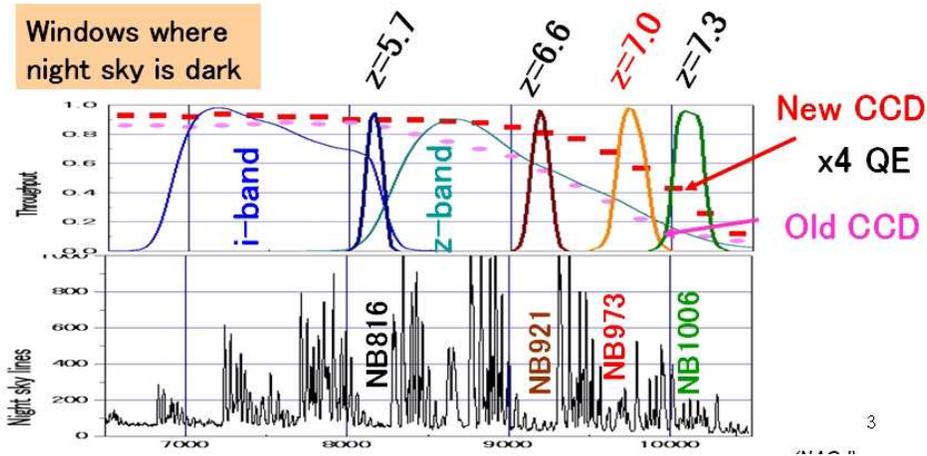}
   \end{tabular}
   \end{center}
  \caption[fig2]
{\label{fig:fig2}
OH night sky emission bands (lower panel) show a few gaps, which astronomers use as dark windows to study deep into the Universe.   Narrow band filters whose transmission are matched to these dark windows are used to sample LAEs at $z=5.7$ (NB816), $z=6.6$ (NB921) and $z=7.0$ (NB973).   The current CCD sensitivity falls rapidly toward 1000nm but recently developed high-resistivity, red-sensitive CCDs  open a possibility to extend the accessible redshift limit up to $z=7.3$. }
\end{figure}

Astronomers use narrow band filters whose transmittance bands are matched to one of these gaps to pick up light only in this gap to detect LAEs whose redshifted Lyman alpha emission enters in this gap. LAEs at appropriate redshift range are expected to show up brighter in the narrow band image than other broad band images.  The narrow band (NB) survey is therefore trying to slice the universe in a narrow range of redshift. There are several such gaps, for instance, the narrow band filter NB816 that has the central wavelength at 816nm is suitable for isolating LAEs at redshift 5.7, NB921nm for redshift 6.6, etc. The most distant LAE at redshift 7.0 confirmed to date was also discovered using the narrow band imaging survey using a filter centered at 973nm. The sensitivity of current CCDs falls rapidly toward 1 micron but recent advent of red sensitive CCDs with thicker depletion layer will extend this redshift limit slightly up to about 7.3.

Let me talk on our discovery of the most distant galaxy.  The red blob in the left panel of Fig. 3 shows the most distant galaxy, IOK-1$^{18}$.   This LAE was discovered among the 41,533 objects in the Subaru Deep Field through the narrow band filter NB973 for a total of 15 hours with  SuprimeCam$^{58}$. All the objects were cross identified in images taken in other filters and only five photometric candidates for $z=7$ LAEs, which are visible only in this narrow band filter, were isolated (cf. Fig.4).  Astronomers have a privilege to name their newly found objects and we took a liberty of naming them taking the initials of three main contributes to this survey, IOK-1 to IOK-5.

\begin{figure}
  \begin{center}
  \begin{tabular}{c}
    \includegraphics[height=6cm]{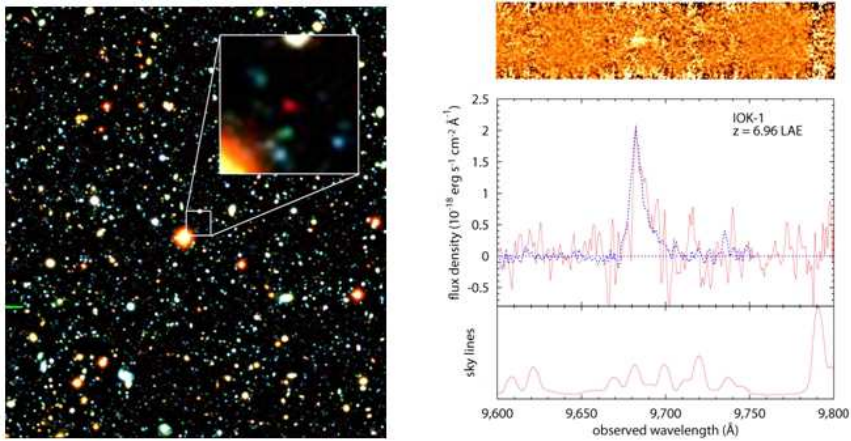}
   \end{tabular}
   \end{center}
  \caption[fig3]
{\label{fig:fig3}
(Left) The most distant galaxy IOK-1 is shown as a red blob in the inlet panel.  (Right)  Spectrum of IOK-1 showing the characteristic Lyman alpha emission line with an asymmetric profile at 968nm indicating its redshift 6.96 (Right panel reproduced from Iye et al., 2006$^{18}$).
}
\end{figure}

\begin{figure}
  \begin{center}
  \begin{tabular}{c}
    \includegraphics[height=2.5cm]{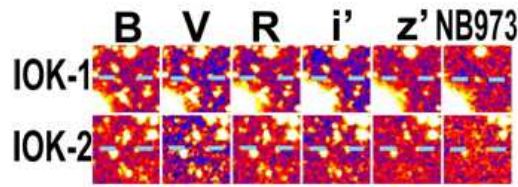}
   \end{tabular}
   \end{center}
  \caption[fig4]
{\label{fig:fig4}
Post stamp images of the NB973 objects IOK-1 and IOK-2.  The latter was confirmed to be a $5\sigma$ noise (Edited from Ota et al., 2008$^{22}$).}
\end{figure}

We have to be, however, careful as there are several types of possible contaminants in these $5\sigma$ photometric candidates. First, since the narrow band imaging observation was made 1-2 year after other broad band observations, some of the candidates may well be variable objects like AGNs or galaxies where supernovae added extra light when narrow band observation was made.  Possibility for emission line objects at lower redshift is a common concern. To our surprise, simple statistics cautions us that there might be one or two $5\sigma$ noises as well, since there are millions of independent 2 arcsec  apertures one can sample in the SuprimeCam field.  Spectroscopic follow-up revealed that only one object, the brightest IOK-1, is a real LAE at redshift 6.96, with the characteristic asymmetric line profile as shown in the right panel of Fig.3.

Table 1 shows the top 10 list of high redshift galaxies with spectroscopic redshift measurement, to the best of my knowledge.  You may notice that 9 out of 10 were discovered by Subaru/SuprimeCam survey in the single Subaru Deep Field.  This is because Subaru/SuprimeCam enables observation of large survey volume with significant depth.   Hubble Ultra Deep Field imaging survey with ACS probes much deeper than ground based observations, but has a much smaller survey volume.  The wide field surveys to pick up scarce bright population and narrow field deep surveys to study fainter populations, are complementary to each other.

\begin{table}
  \caption{Top 10 list of the most distant galaxies with measured redshift.(as of June 6, 2008).}
    \begin{tabular}{lllllll}
\hline
Rank & ID & Coordinates &$z$&Gyr$^{\#}$& Paper & Published date\\
\hline\hline
1 & IOK-1 & J132359.8+272456 & 6.964 & 12.88 & Iye et al. & Sep. 14, 2006\\
2& SDF ID1004 & J132522.3+273520 & 6.597　& 12.82 & Taniguchi et al. & Feb, 25, 2005\\
3& SDF ID1018 & J132520.4+273459 & 6.596　& 12.82 & Kashikawa et al. & Apr. 25, 2006\\
4& SDF ID1030 & J132357.1+272448 & 6.589　& 12.82 & Kashikawa et al. & Apr. 25, 2006\\
5& SDF ID1007 & J132432.5+271647 & 6.580　& 12.82 & Taniguchi et al. & Feb. 25, 2005\\
6& SDF ID1008 & J132518.8+273043 & 6.578　& 12.82 & Taniguchi et al. & Feb. 25, 2005\\
6& SDF ID1001 & J132418.3+271455 & 6.578　& 12.82 & Kodaira et al. & Apr. 25, 2003\\
8$^{*}$& HCM-6A & J023954.7-013332 & 6.560　& 12.82 & Hu et al. & Apr. 1, 2002\\
9& SDF ID1059 & J132432.9+273124 & 6.557　& 12.82 & Kashikawa et al. & Apr. 25, 2006\\
10& SDF ID1003 & J132408.3+271543 & 6.554　& 12.82 & Taniguchi et al. & Feb.25, 2005\\
\hline
 \end{tabular}\\
$^{\#}$ Age in Gyr calculated for a model of the Universe that has an age 13.66 billion years. \\
$^{*}$ This object was discovered by Keck telescope.  All the rest were discovered by Subaru Telescope \\
 in the Subaru Deep Field. \\
   
\end{table}

Subaru Deep Field surveys yielded several dozens of LAE candidates both at redshift 5.7 and 6.6 and about half of them are already confirmed spectroscopically to be LAEs. With this fair sample, one can derive the luminosity function of LAEs. The left panel of Fig.5 shows the UV continuum luminosity functions of LAEs at redshift 5.7 and 6.6 which are, more or less, identical.  On the other hand, the right panel shows the Lyman alpha luminosity functions.  We can see that the brighter population of LAEs at redshift 6.6 is significantly less abundant as compared to those at redshift 5.7.

\begin{figure}
  \begin{center}
  \begin{tabular}{c}
    \includegraphics[height=7.5cm]{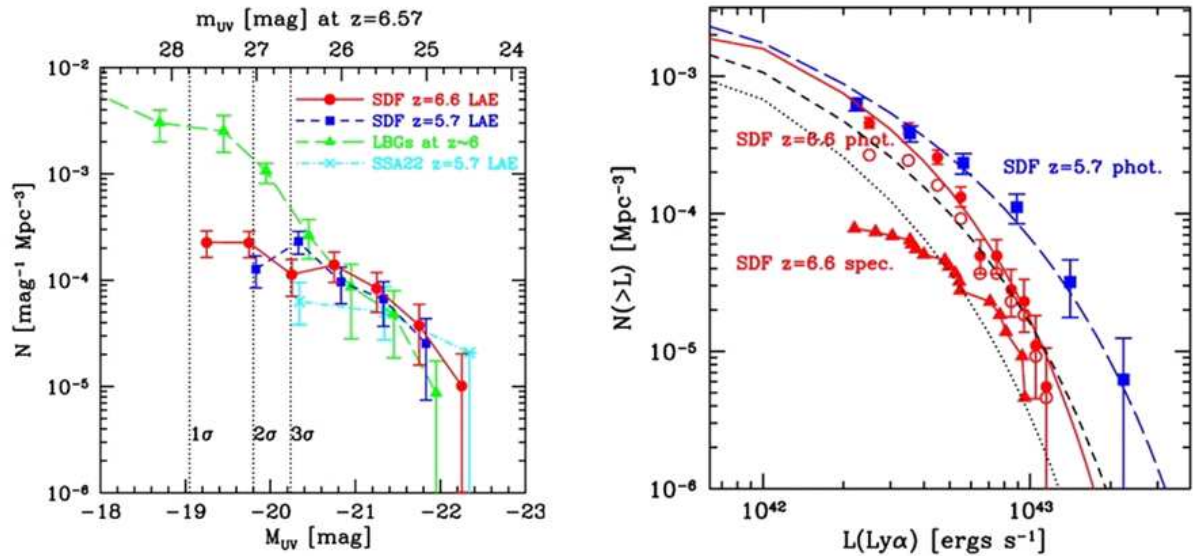}
   \end{tabular}
   \end{center}
  \caption[fig5]
{\label{fig:fig5}
(Left panel) UV continuum luminosity function of LAEs at $z=5.7$ (blue) and $z=6.6$ (red) which are more or less identical.  (Right panel) Lyman $\alpha$ luminosity functions of LAEs at $z=5.7$ (blue) and $z=6.6$ (red).  Note that the significant decrease in Lyaman-$\alpha$ luminosity function at its bright end (Edited from Kashikawa et al., 2006$^{17}$). }
\end{figure}

This can be explained if the neutral hydrogen fraction of the intergalactic matter is increasing from redshift 5.7 to 6.6, as the neutral hydrogen selectively absorbs and scatters the Lyman alpha photons but not for UV continuum. The Lyaman-$\alpha$ luminosity functions, the UV luminosity functions, and the distribution of equivalent width of the LAEs can be reconciled with the presence of Pop III massive star formation followed by PoP II star formation to power Lyaman-$\alpha$ emission$^{60}$.  Of course, the scarcity in LAEs at high redshift could also be due to the evolutionary history of those galaxies building from tiny proto galaxies. Cosmic variance could be another factor, if not significant to this level.

In order to identify LAEs at $z>7$, quite a few projects to make narrow band imaging surveys with near infrared cameras are under way or planned$^{23-28}$.    The field of view of infrared cameras is still considerably smaller than that of, e.g., SuprimeCam and the increasing night sky background make the infrared imaging survey very challenging if the LAE luminosity function is further declining from $z=6.6$ to further redshift.

\section{TWO COLOR DIAGNOSIS FOR LYMAN BREAK GALAXIES}
Another population of galaxies searched for in the early Universe is called Lyman Break Galaxies, abbreviated as LBGs. LBGs are thought to be fairly massive galaxies with evolved stellar population. Stellar continuum is much stronger than LAEs.  Lyman alpha emission is less conspicuous as compared with LAEs.  The spectra of these galaxies show characteristic discontinuity at the blue side of Lyman alpha line caused by the intrinsic stellar atmospheric absorption and by the Intergalactic neutral hydrogen absorption.  These galaxies, therefore, are visible at bands redward of Lyman alpha line but are not visible at bands blueward of the Lyman alpha line. One can select out LBG candidates at $z=6$ by $i$-band dropouts, $z=7$ by $z'$-band dropout, and $z=9$ by $J$-band dropouts.

Here again, one have to be careful for possible contaminants. Galactic T-dwarfs dwell in the similar region in two color diagram.  One may be able to reject T-dwarfs by their point source images if the image quality is superb.   Variable objects and $5\sigma$ noises are the common problems for this survey as wall.

Hubble ACS and NICMOS imaging at Hubble Ultra Deep Field and GOODS field was used to identify faint $z$-dropouts at around $z=7.3$ and about 8 candidates were isolated. but similar attempt for $J$-dropout didn't yield a candidate$^{37}$. Another group reported finding of 10 $z$-dropouts and 2 $J$-dropouts$^{46}$.  Unfortunately, many of these objects do not show strong Lyman alpha emission and spectroscopic confirmation of their genuine redshift is difficult.

\section{SURVEY FOR STRONGLY LENSED GALAXIES}
Let me turn to genius survey projects using the gravitational lensing effect of a massive cluster of galaxies to magnify and brighten the background faint galaxies. Cluster of galaxies are largest telescopes in the Universe with diameter about
1Mpc.  They are nice telescopes for astronomers. You do not need to ask for funding agencies for construction budget and you do not need to ask engineers to design and build them. They are in situ and free of charge to use. Of course there are some drawbacks. You cannot point them to your favorite targets. Wavefront aberrations are bazaar. Although the images produced by cluster lensing are peculiarly deformed and enlarged, the largest advantage is the fact some of the lensed images are brightened considerably and when multiply lensed images are available they can be used to check for the consistency of their reconstructed source image.  

Appropriate modeling of the gravitational field of the cluster enables the prediction of the location of critical lines for assumed source redshift slice where the magnification becomes infinity. Observers can look for lensed object along these critical lines and there are in fact several candidate galaxies found in this way$^{39-47}$.  For instance, a survey for strongly lensed LAEs in 9 clusters yielded six candidates$^{44}$.   If any of these candidates are real, the number density of faint population of galaxies is much larger than previously considered and may well explain the necessary amount of re-ionizing source.

Fig.6 shows a promising z-dropout candidate at redshift 7.6 found behind the cluster Abel 1689 recently$^{45}$.   Photometric results indicate better match to a galaxy at $z=7.6$, however, here again the possibility of galaxy at $z=1.7$ is hard to rule out just from imaging.

\begin{figure}
  \begin{center}
  \begin{tabular}{c}
    \includegraphics[height=8cm]{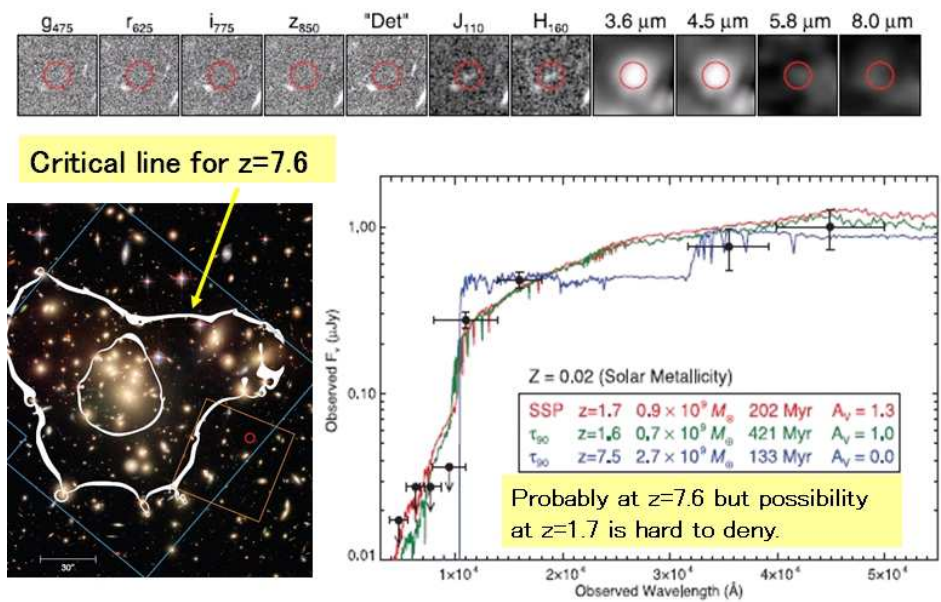}
   \end{tabular}
   \end{center}
  \caption[fig6]
{\label{fig:fig6}
Lyman break galaxy candidate at $z\sim7.6$ discovered behind the lensing cluster A1689 (Edited from Bradley et al. 2008$^{45}$). }
\end{figure}

\section{QUASARS AND GAMMA RAY BURSTERS}
The last objects I am going to introduce are point sources, quasars and gamma ray bursts (GRBs),  in the early Universe. The survey technique used to isolate high redshift quasar candidates is similar to that used for LBGs. Objects that match the expected spectral energy distribution of high redshift quasars are surveyed in the two color diagram or even a multi-dimension color manifold.   Sloan Digital Sky Survey with its enormous data base is a nice test bed to apply this approach. Many quasars beyond redshift 6 were found in this way$^{48-52}$.  The most distant quasar to date is J1148+5251 at 6.42$^{51}$.  Gunn-Peterson test of quasars up to redshift 6 indicated strongly that the cosmic re-ionization ended by redshift 6.

The advent of the real time alert system of gamma ray burst increased the chance of optical and infrared astronomers to make prompt observations of these rapidly declining bursts.  The most distant GRB observed to date is GRB050904 at $z=6.3^{55}$.  GRBs at high redshift can be useful tools to probe the cosmic re-ionization through its Lyman?alpha damping wing$^{56}$.

GRB has a much simpler featureless continuum than the quasar spectra which has broad emission lines superposed on the non-thermal continuum. GRBs are, in a way, better probes to study the re-ionization history.  Both quasars and GRBs are point sources, the advent of laser guide star adaptive optics makes the observation of fainter objects feasible and we expect many such observations if the observatories pay efforts for timely follow-up spectroscopy of long burst GRBs.  GRBs may provide a new way to study even higher-redshift galaxies and first generation of stars.

\begin{figure}
  \begin{center}
  \begin{tabular}{c}
    \includegraphics[height=6cm]{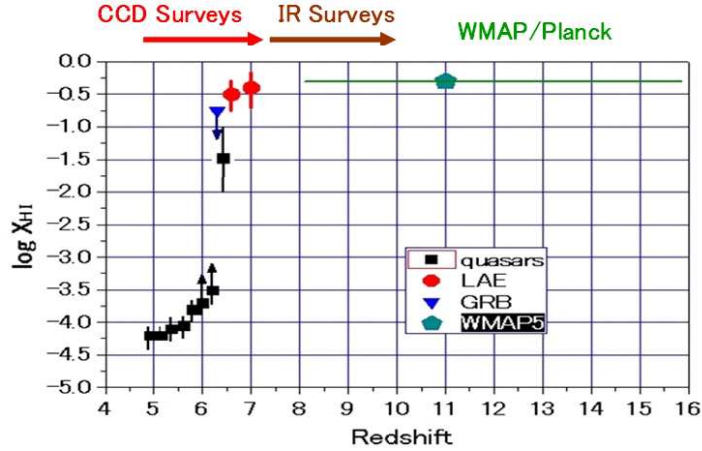}
   \end{tabular}
   \end{center}
  \caption[fig7]
{\label{fig:fig7}
Neutral hydrogen fraction of intergalactic matter as derived from Gunn-Peterson tests of $z>5$ quasars (black squares), damped Lyman-$\alpha$ wing profile (blue triangle), and Lyman alpha luminosity function (red circles).  Also plotted is the WMAP 5 year result, which predict $z=11$ for instantaneous re-ionization.  Note, however, that WMAP cannot constrain when re-ionization started and how long it took to complete.}
\end{figure}

\begin{figure}
  \begin{center}
  \begin{tabular}{c}
    \includegraphics[height=6cm]{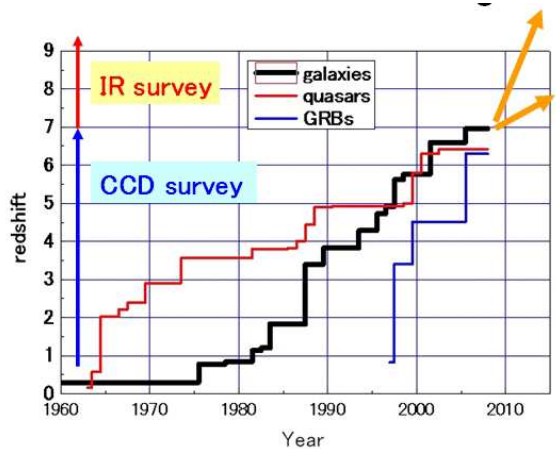}
   \end{tabular}
   \end{center}
  \caption[fig8]
{\label{fig:fig8}
Growth history of largest redshift objects. Note that GRBs are catching up quickly (Based on Tanvir $\&$ Jakobsson, 2007${57}$)}
\end{figure}

Fig.7 shows the increase of the fraction of neutral hydrogen as measured from Gunn-Peterson tests$^{54}$ of quasars up to redshift 6.42 on the left hand. Our results from redshift 6.6 and 7.0 LAE is shown in red and an upper limit from redshift 6.3 GRB is shown in blue triangle. WMAP5 polarization study concludes that the cosmic re-ionization, if it took place instantaneously, would be at redshift around 11$^{59}$  However, WMAP results alone cannot pin down when the cosmic re-ionization started and how log did it take to finish.  Planck satellite my give more clue in 5 years time.  Surveys for galaxies beyond redshift 7 up to 11 is, therefore, extremely important to elucidate what happened actually in this period and for that we need NIR deep surveys.

My last slide (Fig. 8) shows the annual growth of the records of highest redshift objects$^{57}$.  The discovery of our z=6.96 galaxy was announced on Sep.14, 2006, 648 days ago.  Simple statistical argument61 predicts that new record will come in, at 95$\%$ confidence level, at earliest in 17 days from today and at latest in 69 years.  I am confident, however, that we do not need to wait so long as lots of new surveys are under way using near infrared cameras.  Besides, observations of　GRBs are catching up quickly, and considering the availability of innovated LGSAO, I would rather predict GRB will soon take over this race.

\section{REFERENCES}
\small
[1] Partridge, R.B. \& Peebles, P. J. E., "Are young galaxies visible?", Astrophys. J., 147, 868-886 (1967)

\noindent
[2] Pritchet, C.J., "The search for primeval galaxies", Publ. Astron. Soc. Pacific, 106, 1052-1067 (1994)

\noindent
[3] Hu, E. M. \& McMahon, R. G., "Detection of Lyaman-$\alpha$-emitting galaxies at redshift 4.55", Nature 382, 231-233 (1996).

\noindent
[4] Cowie, L. L. \& Hu, E. M., "High-z Ly-alpha emitters. I. A blank-field search for objects near redshift $z = 3.4$ in and around the Hubble Deep Field and the Hawaii Deep Field SSA22",  Astron. J. 115, 1319-1328 (1998).

\noindent
[5] Loeb, A. \& Rybicki, G. B., "Scattered Ly-alpha radiation around sources before cosmological re-ionization", Astrophys. J. 524, 527-535 (1999).

\noindent
[6] Hu, E. M. et al., "A redshift $z = 6.56$ galaxy behind the cluster Abell 370", Astrophys. J. 568, L75-L79 (2002).

\noindent
[7] Kodaira, K. et al., "The discovery of two Lyman alpha emitters beyond redshift 6 in the Subaru Deep Field", Publ. Astron. Soc. Jpn 55, L17-L21 (2003).

\noindent
[8] Santos, M. R., Ellis, R. S., Kneib, J.-P., Richard, J. \& Kuijken, K., "The abundance of low luminosity Ly-alpha emitters at high redshift", Astrophys. J. 606, 683-701 (2004).

\noindent
[9] Hu, E. M. et al., "The luminosity function of Ly-alpha emitters at redshift $z\sim5.7$", Astron. J. 127, 563-575 (2004).

\noindent
[10] Kurk, J. D. et al., “A Lyman alpha emitter at $z = 6.5$ found with slitless spectroscopy", Astron. Astrophys. 422, L13?L17 (2004).

\noindent
[11] Rhoads, J. E. et al., "A luminous Ly-alpha-emitting galaxy at redshift $z = 6.535$: Discovery and spectroscopic confirmation", Astrophys. J. 611, 59-67 (2004).

\noindent
[12] Stanway, E. R. et al., "Three Ly-alpha emitters at $z\sim 6$: Early GMOS/Gemini data from the GLARE project",  Astrophys. J. 604, L13-L16 (2004).

\noindent
[13] Malhotra, S. \& Rhoads, J. E., "Luminosity functions of Ly-alpha emitters at redshifts $z = 6.5$ and $z = 5.7$: Evidence against re-ionization at $z < 6.5$", Astrophys. J. 617, L5-L8 (2004).

\noindent
[14] Nagao, T. et al., "A strong Ly-alpha emitter at $z = 6.33$ in the Subaru Deep Field selected as an $i'$-dropout", Astrophys. J. 613, L9-L12 (2004).

\noindent
[15] Taniguchi, Y. et al., "The SUBARU Deep Field Project: Lyman alpha emitters at a redshift of 6.6",  Publ. Astron. Soc. Jpn 57, 165-182 (2005)

\noindent
[16] Shimasaku, K., et al, "Ly alpha Emitters at $z=5.7$ in the Subaru Deep Field", PASJ, 58, 313-334, (2006)

\noindent
[17] Kashikawa, N., et al., "The End of the Re-ionization Epoch Probed by Ly-alpha Emitters at $z=6.5$ in the Subaru Deep Field", Astrophys.J, 648, 7, (2006)

\noindent
[18] Iye, M. et al., "A Galaxy at a redshift 6.96", Nature, 443, 186-188 (2006)

\noindent
[19] Hu, E.M. \& Cowie, L.L., "High-redshift galaxy populations", Nature, 440, 1145-1150 (2006)   

\noindent
[20] Ouchi, M. et al., "Exploring the Cosmic Dawn with Subaru Telescope", ASP Conf. Series, 379, 47-54 (2007)

\noindent
[21] Stark, D.P. et al., "An empirically calibrated model for interpreting the evolution of galaxies during the re-ionization era", Astrophys.J., 668, 627-642 (2007)

\noindent
[22] Ota, K. et al., "Re-ionization and galaxy evolution probed by $z=7$ Ly-alpha emitters", Astrophys.J., 677, 12-26 (2008)  

\noindent
[23] Horton, et al., "DAzLE: The Dark Ages z(redshift) Lyaman-$\alpha$ Explorer", SPIE, 5492, 1022, (2004)

\noindent
[24] Ichikawa,T. et al., "MOIRCS Deep Survey II. Clustering properties of K-band selected galaxies in GOODS-North region", PASJ, 59, 1081-1094 (2007)

\noindent
[25] Cuby, et al., "A narrow-band search for Ly-alpha emitting galaxies at $z=8.8$", Astron \& Astrophys., 461, 911-916 (2007)

\noindent
[26] Nilsson, K.K. et al., "Narrow-band surveys for very high redshift Lyaman-$\alpha$ emitters", Astron. \& Astrophys. 474, 385-392 (2007)

\noindent
[27] Geach, J.E.,  et al., "HiZELS: a high redshift survey of H-alpha emitters. I: the cosmic star-formation rate and clustering at $z=2.23$", arXiv:0805.2861 (2008)

\noindent
[28] Smail,I. et al., "The HiZELS Survey", http://astro.dur.ac.uk/$^{\sim}$irs/HiZELS/

\noindent
[29] Stanway, E. R., Bunker, A. J. \& McMahon, R. G., “Lyman break galaxies and the star formation rate of the Universe at $z\sim 6$", Mon. Not. R. Astron. Soc. 342, 439-445 (2003).

\noindent
[30] Bouwens, R., Broadhurst, T. \& Illingworth, G., "Cloning dropouts: Implications for galaxy evolution at high redshift",  Astrophys. J. 593, 640-660 (2003).

\noindent
[31] Bouwens, R. J. et al., "Galaxies at $z\sim7-8$: $z_{850}$-dropouts in the Hubble Ultra Deep Field", Astrophys. J. 616, L79-L82 (2004).

\noindent
[32] Bunker, A. J., Stanway, E. R., Ellis, R. S. $\&$ McMahon, R. G., “The star formation rate of the Universe at z$\sim$ 6 from the Hubble Ultra-Deep Field", Mon. Not. R. Astron. Soc. 355, 374-384 (2004).

\noindent
[33] Stanway, E. R. et al., "Hubble Space Telescope imaging and Keck spectroscopy of z$\sim6$ $i$-band dropout galaxies in the Advanced Camera for Surveys GOODS fields", Astrophys. J. 607, 704-720 (2004).

\noindent
[34] Yan, H. $\&$ Windhorst, R. A., "Candidates of$ z\sim 5.5-7$ galaxies in the Hubble Space Telescope Ultra Deep Field", Astrophys. J. 612, L93-L96 (2004).

\noindent
[35] Bouwens, R. $\&$ Illingworth, G., "Rapid Evolution in the most luminous galaxies during the first 900 million years", Nature 443, 189-192 (2006)

\noindent
[36] Henry, A.L., et al., "A Lyman break galaxy candidate at $z\sim9$", arXiv:0805.1228v1 (2008),  NICMOS/IRAC photometric survey for $J$ dropouts.

\noindent
[37] Bouwens, R. et al., "$z\sim7-10$ galaxies in the HUDF and GOODS fields, and their UV luminosity functions", arXiv-0803.0548v2 (2008) 

\noindent
[38] Oesch, P. A., "The UDF05 follow-up of the HUDF: II. Constraints on re-ionization from $z$-dropout galaxies", arXiv:0804.4874v1 (2008)    

\noindent
[39] Kneib, J.-P., Ellis, R. S., Santos, M. R. $\&$ Richard, J., "A probable $z\sim7$ galaxy strongly lensed by the rich cluster A2218: Exploring the Dark Ages", Astrophys. J. 607, 697-703 (2004).

\noindent
[40] Egami, E. et al., "Spitzer and Hubble Space Telescope constraints on the physical properties of the $z\sim7$ galaxy strongly lensed by A2218", Astrophys. J. 618, L5-L8 (2005).   A galaxy lensed by A2218 possibly at redshift $z\sim6.6-6.8$.

\noindent
[41] Willis, J. P. $\&$ Courbin, F., "A deep, narrow J-band search for protogalactic Ly-alpha emission at redshifts $z\sim9$", Mon. Not. R. Astron. Soc. 357, 1348-1356 (2005).

\noindent
[42] Yan, H. et al., "Rest-frame ultraviolet-to-optical properties of galaxies at $z\sim6$ and $z\sim5$ in the Hubble Ultra Deep Field: From Hubble to Spitzer", Astrophys. J. 634, 109-127 (2005)

\noindent
[43] Chary, R.-R., Stern, D. $\&$ Eisenhardt, P., "Spitzer constraints on the $z = 6.56$ galaxy lensed by Abell 370", Astrophys. J. 635, L5-L8 (2005)

\noindent
[44] Stark, D.P. et al., "A Keck survey for gravitationally-lensed Lyaman-$\alpha$ emitters in the redshift range $8.5 < z < 10.4$: New constraints on the contribution of low luminosity sources to cosmic re-ionization", Astrophys.J. 663, 10-28 (2007)   

\noindent
[45] Bradley, L.D., et al., "Discovery of a very bright strongly lensed galaxy candidate at $z\sim7.6$", Astrophys. J. 678, 647-654 (2008)

\noindent
[46] Richard, J. et al., "A Hubble $\&$ Spitzer Space Telescope Survey of Gravitationally-lensed galaxies: Further evidence for a significant population of low luminosity galaxies beyond redshift seven", arXis:0803.4391v2  

\noindent
[47] Yan, H.J., et al.,  "Search for Very High-z Galaxies with WFC3 Pure Parallel / HST Proposal 11702", http://archive.stsci.edu/cgi-bin/proposal$\_$search?mission=hst$\&$id=11702

\noindent
[48] Schneider, D. P., Schmidt, M. $\&$ Gunn, J. E., "PC 1247 + 3406: An optically selected quasar with a redshift of 4.897", Astron. J. 102, 837-840 (1991).

\noindent
[49] Fan, X. et al., "A survey of $z > 5.8$ quasars in the Sloan Digital Sky Survey. I. Discovery of three new quasars and the spatial density of luminous quasars at $z\sim6$", Astron. J. 122, 2833-2849 (2001).

\noindent
[50] Becker, R. H. et al., "Evidence for re-ionization at $z\sim6$: Detection of a Gunn-Peterson trough in a $z = 6.28$ quasar",  Astron. J. 122, 2850-2857 (2001).

\noindent
[51] Fan, X. et al., "Evolution of the ionizing background and the epoch of re-ionization from the spectra of $z\sim 6$ quasars", Astron. J. 123, 1247-1257 (2002).

\noindent
[52] Fan, X. et al., "A survey of $z > 5.7$ quasars in the Sloan Digital Sky Survey. II. Discovery of three additional quasars at $z > 6$",  Astron. J. 125, 1649-1659 (2003).

\noindent
[53] Fan, X. et al., "Constraining the evolution of the ionizing background and the epoch of re-ionization with $z\sim 6$ quasars II: A sample of 19 quasars", ArXiv Astrophysics e-prints $<$astro-ph/0512082$>$ (2006).

\noindent
[54] Gunn, J. E. $\&$ Peterson, B. A., "On the density of neutral hydrogen in intergalactic space", Astrophys. J. 142, 1633-1641 (1965).

\noindent
[55] Kawai, N. et al., "Afterglow spectrum of a gamma-ray burst with the highest known redshift $z = 6.295$", Nature, 440, 184-186 (2006).

\noindent
[56] Totani, T. et al., "Implications for cosmic re-ionization from optical afterglow spectrum of the Gamma-Ray Burst 050904 at $z = 6.3$", PASJ, 58, 485, (2006)

\noindent
[57] Tanvir, N. R. $\&$ Jakobsson, P., "Observations of GRBs at high redshift", astro-ph/0701777 (2007)

\noindent
[58] Miyazaki, S. et al., "Subaru prime focus camera: Suprime-Cam", Publ. Astron. Soc. Jpn 54, 833-853 (2002).

\noindent
[59] Dunkley, J. et al., "Five-year Wilkinson Microwave Anisotropy Probe (WMAP) Observations: Likelihoods and Parameters from the WMAP data", arXiv:0803.0586v1, (2008)   

\noindent
[60] Dijkstra,M. and Wyithe, S.B., "Very massive star in high-redshift galaxies", Mon. Not. R. Astron. Soc., 379, 1589-1598 (2007) 

\noindent
[61] Gott, III, J.R., "Implications of the Copernican principle for our future prospects", Nature, 363, 315-319 (1993)

\bibliography{report}   
\bibliographystyle{spiebib}   
\end{document}